\documentstyle[preprint,aps,prb]{revtex}
\begin{document}
\draft
\title{Topologically induced vortex magnetance of a quantum device}
\author{P.~Exner, P.~\v{S}eba and A.F.~Sadreev$^{\dag}$,}
\address{
Nuclear Physics Institute, Academy of Sciences,
250 68 \v{R}e\v{z} near Prague \\ and \\
Doppler Institute, Czech Technical University,
B\v{r}ehov\'{a} 7, 115 19 Prague, Czech Republic}
\author{P.~St\v{r}eda and P.~Feher}
\address{
Institute of Physics, Academy of Sciences, Cukrovarnick\'{a}
10, 162 00 Prague, Czech Republic}

\date{\today}

\maketitle

\begin{abstract}
We consider resonant vortices around nodal points of
the wavefunction in electron transport through a mesoscopic device.
With a suitable choice of the device geometry, the dominating role
is played by  single vortices of a preferred orientation.
To characterize strength of the resulting magnetic moment we
have introduced a magnetance, the quantity defined in analogy
with the device conductance. Its basic properties and possible
experimental detection are discussed.
\end{abstract}


\narrowtext
The topological structure of quantum mechanical wavefunctions is
responsible for many observable phenomena. One of its prominent
consequences is  existence of vortices. They have already
been observed in macroscopic quantum systems, rotating superfluid
helium and superconductors.
In general, their existence is connected with the time reversal
symmetry breaking. The first example of the such topological effect
was already given in the very early days of the quantum theory.
It concerns the probability current density of a single electron
defined conventionally by
\begin{equation}
\vec j(\vec r)=
(\hbar/m^*)\, {\rm Im\,} \left(\bar\psi(\vec r)\vec\nabla\psi(\vec
r)\right)
\label{current}
\end{equation}
which may be nontrivial once the wavefunction
$\,\psi\,$ is complex. In systems with open geometry a nonzero
probability current refers to an electron transport
through the system, i. e. from the source to a drain.
Dirac \cite{Dirac} pointed out that the quantum probability
current may exhibit a vortex structure around nodal points (zeros) of
the corresponding wavefunction. The effect was later discussed by
Hirschfelder \cite{Hirsch}. Recently it has been demonstrated
numerically that the current in topologically nontrivial devices
exhibits pronounced vortices the form and magnitude of which changes
quickly with the energy \cite{BBJ,CBC,ESTV,RVZ,WS} and applied
magnetic field \cite{PS}. The experimental evidence of their
existence is of the particular interest.

The direct consequence of a vortex is a non-zero magnetic moment,
which has to depend on the applied current in a similar way as
the device conductance. The natural physical quantity representing
a vortex structure strength is thus the device magnetance given
by the ratio of the magnetic moment and applied voltage drop.
To analyse the effect of vortices to the electron transport and
magnetance we will limit our consideration to devices with
two-dimensional gas of spinless electrons. Main attention will be
payed to the simple device formed by a piece of the
one--dimensional wire with
tangentially attached circular cavity (a quantum dot) as sketched
in Fig.~1. It may be expected that there appears a dominating
circulating current in analogy with the classical gas flow.
This structure is thus the natural candidate for an experimental
attempt to observe electron vortices.

The vortices are closely related with  the topological structure of
the phase. To be more explicit, let us write the
wavefunction as
\begin{equation}
\psi=\sqrt{\rho}\, e^{iS}\,,
\label{phase}
\end{equation}
where $\,\rho\,$ is the probability density of the particle and
$\,S\,$ is the corresponding phase.  The latter is defined modulo
$\,2\pi\,$ and assumes conventionally values in the interval
$\,[0,2\pi)\,$. In the two--dimensional case it is possible to identify
the configuration space with a region in the complex plane and to
treat a possible multivaluedness of $\,S\,$ in terms of its
analytical structure, especially cuts and branching points
corresponding to different Riemannian sheets of this function.

Existence of a branching point on the phase Riemannian surface
implies appearance of a vortex in the corresponding quantum
probability flow \cite{Hirsch}. Inserting (\ref{phase}) into
(\ref{current}) we get
\begin{equation}
\vec j(\vec r)\,=\,
\frac{\hbar}{m^{\ast}} \,  \rho(\vec r)\vec\nabla S(\vec r)\,.
\label{current2}
\end{equation}
The vector $\,\vec v=(\hbar/m^*)\vec\nabla S\,$ can be regarded as a
velocity of the corresponding probability flow. For a closed
curve $\Gamma$ the vorticity of $\,\vec v\,$ along $\,\Gamma\,$ is
thus
\begin{equation}
\oint_\Gamma \vec v~d\vec l\,=\, \frac{\hbar}{m^{\ast}}\,\delta S\,
= \, 2\pi \frac{\hbar}{m^{\ast}} \, m \; ; \; m=0,\pm 1, \pm
2,\dots\,,
\label{vorticity}
\end{equation}
where $\,\delta S\,$ is the phase change when winding once around the
curve. Since the wave function (\ref{phase}) must be single--valued, the
difference $\,\delta S\,$ can only equal to a multiple of $\,2\pi\,$.
If the phase $\,S\,$ has no singularities inside $\,\Gamma$, the
contour can be shrinked into a point in which case the vorticity
is zero. If, on the other hand,
$\Gamma$ encircles a nodal point of $\,\psi$, the phase is ambiguous
at this point and the integer $\,m\,$  may be
nonzero. In this situation the corresponding probability current
exhibits a vortex centered at the nodal point.

In quantum devices three basic topological situations can occur
(see Fig.~1)~:
\\
(a) A~phase cut starts and ends at the boundary of the system. This
is typical for integrable systems. As a simple example, consider the
transport through a straight quantum waveguide of the width $\,w \,$
parallel to the $\,x$--axis. If the incident wave is in the $\,n$--th
transverse mode, the wavefunction has the form
\begin{equation}
\psi_{k,n}(x,y) \, = \, \sqrt{\frac{1}{\pi w}} \,e^{ikx}\,
\, \sin \frac{\pi n y}{w} \; \; \; \; ,
\label{transmode}
\end{equation}
where $\,n\,$ is the mode number and $\hbar^2 (\,k^2\!+ \pi^2
n^2/w^2 )/2 m^{\ast}$ is the particle
energy. In this case the phase $\,S=kx\,$ is monotonically increasing
in the longitudinal direction and its cuts are located at the
segments $\,x=2\pi j/k\,,\;j=\dots,-1,0,1,\dots\,$, parallel to the
$\,x$--axis. Their endpoints lie at the boundary, $\,y=0, w\,$.
They are physically irrelevant being not branching points. \\
(b) The cut starts at the boundary and ends at some nodal point
inside the system. The internal endpoint is a branching point which
is related to a single vortex. \\
(c) The cut connects two nodal points of the wavefunction. Both
endpoints are branching points which corresponds to a pair of
vortices rotating in opposite directions.

To define the vortex magnetance we will follow the
scattering approach \cite{Fisher}, generally
accepted in the transport theory of quantum devices. To establish
conductance of a two terminal device, infinitely long ideal leads
are placed between the device (scattering region) and
electron reservoirs (source and drain) allowing an explicit
asymptotic form of scattering wave functions. The obtained
transmission coefficient $T(E)$ is used to relate the
applied current $\cal{J}$ and the chemical potential difference
between reservoirs, $\Delta \mu$.
In the limiting case of vanishing $\Delta \mu$ and at
zero temperature the electron transport is determined by
properties of electrons at the Fermi energy $E_F$ and we have
\begin{equation}
{\cal J} \; = \; \frac{e}{h} \, T(E_F)  \, \Delta \mu \;.
\label{transmise}
\end{equation}

The applied current is responsible for a non-zero  momentum of the
system. It can be divided into two parts: momentum of the
center-of-mass and the momentum due to electron motion relative
to the center-of-mass. The later one originates in circulating
currents giving rise to a magnetic moment
$\, \vec{M} \equiv (0,0,M)$ representing their strength.
Momentum of the center-of-mass is controlled by the current
density within the ideal leads only.
Note that in the considered limit of infinitely long ideal leads
the mass center momentum cannot be affected by a device of finite
dimensions.

In the simplest case of one--dimensional ideal leads having a
common axis the momentum of the center-of-mass can easily be
evaluated. It is controlled by the
product $\, T(E_{F})\vec{j}_{0} (\vec{r}) \,$ with
$\, \vec{j}_{0} (\vec{r}) \,$ being the current density
within the system without the scattering region
(the device is replaced by the lead). In this case
we get the vortex magnetance in the following form:
\begin{equation}
\frac{\vec{M}(E_F)}{\Delta \mu}  =  \frac{e g(E_F)}{2c}
\int \left (  \vec{j}(\vec{r}) - T(E_F) \vec{j}_0 (\vec{r})  \right )
\times  \vec{r} \, d^2 r   ,
\label{moment}
\end{equation}
where $g(E)$ denotes the density of device states.
The magnetance defined by this way characterizes the vortex
structure of the studied device and it is invariant with
respect to  coordinate system translations.

The expression, Eq.(\ref{moment}), is applicable to the device
sketched in Fig.~1. To stress the effect of the device
geometry the flat potential
is assumed within the device area demarcated by hard walls.
The relevant wave functions are thus eigenfunctions  of the
Schr\"{o}dinger equation for free electrons with zero values
at the boundaries of the device and ideal leads.

To estimate the magnetance of the considered device,
let us first summarize
wavefunction properties of the separated circular cavity.
In polar coordinates eigenfunctions  are determined by zeros
of Bessel functions $J_m$ at the cavity radius $R$ and we have
\begin{equation}
\psi_{l,m}^{\pm}(r,\theta) \,=\, c_{l,m} \, J_m(k_{l,m}  r)
e^{\pm i m \theta}
\; \; \; ,
\label{wave}
\end{equation}
where $\,k_{l,m}=x_{l,m}/ R \,$ with $\,x_{l,m}$ being the
$\, l$--th zero of the Bessel function $\,J_m$, and $\, c_{l,m} \,$
is the normalization factor. The quantum number $m$
($\, m = 0,\, 1, \, 2, \, \ldots \,$) represents the angular momentum
$\pm(\hbar/m^{\ast})m$ which is closely related to the vorticity
defined by Eq.(\ref{vorticity}).
The eigenstates of spinless electrons are
two--fold degenerated with respect of the sign of the momentum.

Only the azimuthal component of the probability current density for
given eigenenergy $E_{l,m}$ might be nonzero and it is given as follows
\begin{equation}
j_{\theta}(r) \, = \, A_{l,m} \, \frac{\hbar}{m^{\ast}} \,
\frac{m}{r} \,c_{l,m}^{2} \, J_m^2(k_{l,m}r) \; \; \; ,
\label{j_theta}
\end{equation}
where $A_{l,m}$ is given by the difference of weight factors,
$\, A_{l,m} \, = \, |a_{l,m}^{(+)}|^2 - |a_{l,m}^{(-)}|^2 \,$,
representing amplitudes $\, a_{l,m}^{(+)} \, $ and
$\, a_{l,m}^{(-)} \,$ of eigenfunctions with positive and
negative values of the orbital momentum, respectively.
In the equilibrium case the problem has time reversal
symmetry requiring equality of weight factors and the total
momentum vanishes.

Attaching a wire the wavefunctions $\, \psi(r,\theta)\,$
given by Eq.(\ref{wave}) become modified by a coupling with
plane waves. Assuming a small window between wire and cavity,
much less than R,
wave functions $\psi(r,\theta)$ will only be slightly
modified. The main effect will be a level broadening
represented by the spectral density $g_{l,m}(E)$ with
sharp maxima at energies $E \cong E_{l,m}$.
Wave functions representing states with the same direction
of the probability flow in the window region will be much
easily matched together than those with opposite flow
directions. In the current currying regime the
amplitude of cavity wave functions matched with incident
plane wave will be thus enhanced giving rise to circulating
currents. Corresponding magnetance may be estimated by
choosing a proper value of the $A_{l,m}(E)$
entering Eq.(\ref{j_theta}). Inserting the current density
$j_{\theta}(r)$, Eq.(\ref{j_theta}), into expression for the
magnetance, Eq.(\ref{moment}), we get
\begin{equation}
\frac{M(E_F)}{\Delta \mu} \; \cong \; \mu_B^{\ast} \,  \sum_{l,m}
A_{l,m}(E_F) \, m \, g_{l,m}(E_F) \,   \; \; \; ,
\label{qmagnetance}
\end{equation}
where $\mu_{B}^{\ast}=e\hbar/2m^{\ast}c$ is the effective Bohr
magneton. Due to the dominating role of the spectral density
$g_{l,m}(E)$ a resonance character of the magnetance is expected.
Since  $g_{l,m}(E)$ is proportional to $m^{\ast}$ the magnetance is
of the purely topological origin.

As an example, the energy dependence of the magnetance
and the transmission coefficient close to the resonance for
$l=1$ and $m=3$ is shown in Fig.~2. Corresponding current density
distribution for $E_F$ indicated by arrow is presented in Fig.~1.
Four phase-cut endpoints represent vortex centers. Three of them
are located close to the cavity center giving rise to pronounced
circulating current. For fixed window
region the broadening of cavity levels becomes weaker for larger
radius $R$ giving rise to more pronounced resonances as
seen in Fig.~3a. For $R > 1.8 \, w$
more complicated resonance structure appears due to states with
different quantum numbers $l$.

For a quite well separated resonance the magnetance peak height
may be estimated by equality of the current originated  in the
cavity eigenfunction, Eq.(\ref{wave}), and that  of the single
wire, Eq.(\ref{transmode}), in the common region, i.e. in the
area of overlapping circular and strip regions.
This condition gives:
\begin{eqnarray}
A_{l,m}(E_r) \frac{\hbar g(E_r)}{m^{\ast}} \, m \, \int_{R-d}^{R}
c_{l,m}^{2} J_m^2(k_{l,m}  r) \frac{dr}{r} \; \cong \;
\nonumber  \\
\cong \; \frac{1}{h}  \, \frac{1}{\pi w}  \,
\int_{w-d}^{w} \sin^{2} \frac{\pi y}{w} \, dy \; \; \; ,
\label{condition}
\end{eqnarray}
where $d$ is the width of the common region, $d = w + R - y_{0}$
with $y_{0}$ being the distance of the cavity center from the
bottom wire edge. This estimation is in good agreement with the
values obtained by direct evaluation of the Eq.(\ref{moment}),
as shown in Fig.~3a.

The above amplitude estimation, Eq.(\ref{condition}), assumes
an {\it ideal} coupling allowing an easy transfer of the incident
wave with transmission coefficient $T$ approaching
unity. Out of the {\it ideal} coupling a reflection
will take place and $T$ will decrease. However, far enough from
the resonance  the mixing of plane waves with eigenfunction of
the separated cavity becomes less effective and transmission
will be enhanced. This typical energy dependence is shown in
Fig.~2b.

At finite temperatures the magnetance is given by the following
expression
\begin{equation}
\frac{M(\mu,T)}{\Delta \mu} \; = \; - \, \frac{1}{\Delta \mu} \,
\int \frac{\partial f_{0}(E)}{\partial \mu}
\, M(E) \, dE \; \; \; ,
\label{avmoment}
\end{equation}
where $f_{0}(E)$ denotes the Fermi--Dirac distribution function
and $M(E)$ is the zero temperature magnetic moment given by
Eq.(\ref{moment}). The resulting magnetance at a non-zero
temperature  is shown in Fig.~3b. For high enough
temperatures  resonances will fully be smeared out and the magnetance
should approach values obtained by a semiclassical treatment.
Since the averaged density of states $g$ scales with the cavity
area $<g> \cong (m^{\ast}/h \hbar) E_{F} \, \pi R^{2}$
and $m \sim R/w$, the approximate scaling of the magnetance
with $R^{3}$ is expected.

The semiclassical description of the studied ballistic transport
can be based on the billiard model \cite{biliar}. Electrons
are injected uniformly over the wire width $w$ and with the
probability $\frac{1}{2} \cos \phi$ along the direction
represented by the angle $\phi$ with respect of the wire axis.
As expected, the resulting
magnetance, shown in Fig.~3b, approaches mean values obtained
by full quantum--mechanical treatment.

The same qualitative features of the magnetance are expected for
any cavity defined by a potential of the circular symmetry. We
have found that resonances are very stable with respect of the
radius modulation ($\Delta R(\theta)/\langle R \rangle <$0.1) and
potential fluctuations less than the energy level separation of
the unperturbed cavity. Interaction between electrons, which was not
taken into account, should not change the magnetance structure
qualitatively. However, we expect enhanced magnetance due to the
electron-electron friction.

Impurity potentials within wire area have destructive effect to
the magnetance. They suppress the transmission coefficient and
consequently the current flow along window between wire and
cavity, which is responsible for the magnetance peak height as
discussed above, Eq.(\ref{condition}). They also give rise to
additional vortices of the uncontrollable orientation within
the wire region. The resulting richer magnetance structure can
be much easily smeared out by the temperature. This destructive
effects will be even more effective for wider, multimode, wires.

In real systems the probability to have impurities within wire
region might be lowered by the shortening of the wire length.
Similar device, but with centrally
attached wires represented by point contacts, have already been
realized \cite{Wees}.  The proper design could also allow to
control cavity area by  top gates. To observe resonances the
energy level separation of the isolated cavity has to be larger
than the thermal energy $k_B T$. For the device fabricated from
a GaAs-AlGaAs heterostructure it limits the cavity radius,
$R[\mu$m$] < 0.2/\sqrt{T[{\rm K}]}$.
For example, assuming $R \sim 0.5 \mu$m resonances could be
indicated by a fluxmeter allowing measurements with accuracy
10$^{2}$ $\mu_B$, which is in principle within experimental reach
\cite{Geim}. The contribution of non-circulating currents cannot
be easily excluded by experimental setup as has been done in our
theoretical treatment. Under the standard
conditions of the fixed applied current it will be, however,
responsible for a monotonic background only.

\acknowledgments
The research has been partially supported by the DTP
Foundation as well as by the Grants INTAS-RFBR 95--657 and GACR
202--0218/96--98. The support obtained from the Theoretical
Physics Foundation in Slemeno is also greatly acknowledged.

\begin{figure}
\caption{The model device geometry with phase cuts (solid lines)
for the energy indicated in Fig.~2 by the arrow. The window
between wire and the cavity equals to 5/4 of the wire width $w$
and the cavity radius $R=5w/3$. Corresponding current density
is illustrated by arrows.}
\label{fig1}
\end{figure}

\begin{figure}
\caption{Energy dependence of the magnetance $M/\Delta \mu$ (a)
and transmission coefficient $T$ (b) in the vicinity of the
resonance for $l=1$ and $m=3$. Device geometry is the same as
that in Fig.~1 and $E_1$ denotes the energy of the lowest
transversal mode $E_1=(\pi^2 \hbar^2)/(2 m^{\ast} w^2)$.}
\label{fig2}
\end{figure}

\begin{figure}
\caption{Magnetance $M/\Delta \mu$ at zero temperature (a) and
at the temperature $0.05 \, E_1 /k_B$ (b) as function of the
cavity radius for fixed window width $5w/4$. The dashed line
represents estimation for resonance maxima, Eq.(11). The
dashed-dotted line is the result of the semiclassical treatment.}
\label{fig3}
\end{figure}


\begin{references}

\bibitem[{\dag}] ppermanent address: Institute of Physics, Academy of Sciences,
660036 Krasnoyarsk, Russia

\bibitem{Dirac}
P.A.M.~Dirac, Proc. Roy. Soc. {\bf A~133}, 60 (1931).

\bibitem{Hirsch}
J.O.~Hirschfelder,  J. Chem. Phys. {\bf 67}, 5477 (1977).

\bibitem{BBJ}
K.--F.~Berggren, and Zheng--Li Ji, Phys. Rev.
{\bf B 47}, 6390 (1993).

\bibitem{CBC}
S.~Chaudhuri, S.~Badyopadhyay, and M.~Cahay, Phys. Rev.
{\bf B 45}, 11126 (1992).

\bibitem{ESTV}
P~Exner, P.~\v Seba, M.~Tater, and D.~Van\v ek, J. Math. Phys.
{\bf 37}, 4867 (1996).

\bibitem{RVZ}
V.M.~Ramaglia, F.~Ventriglia, and G.P.~Zucchelli, Phys. Rev. {\bf
B 48}, 2445 (1993); {\bf B 52}, 8372 (1993).

\bibitem{WS}
Hua Wu, D.W.L.~Sprung, {\em Phys. Lett.} {\bf A183}, 413 (1993).

\bibitem{PS}
K.N.Pichugin, and A.F.Sadreev, JETP {\bf 82}, 290 (1996).

\bibitem{Fisher}
D.S.~Fisher, and P.A.~Lee, Phys. Rev. {\bf B 23}, 6851 (1981).

\bibitem{biliar}
C.W.~Beenakker, and H.~van~Houten, Phys. Rev. Lett.
{\bf 63}, 1857 (1989).

\bibitem{Wees}
B.J.~van~Wees, L.P.~Kouwenhoven, C.J.P.M.~Harmans,
G.J.~Williamson, C.E.Timmering, M.E.I.~Broekaart, C.T.~Foxon, and
J.J.~Harris, Phys. Rev. Lett. {\bf 62}, 2523 (1989).


\bibitem{Geim}
A.K.~Geim, S.V.~Dubonos, I.V.~Grigorieva, J.G.S.~Lok, J.C.~Maan,
X.Q.~Li, F.M.~Peeters, and Yu.V.~Nazarov, {\em will appear in}
Superlattices and Microstructures (1997).


\end{references}
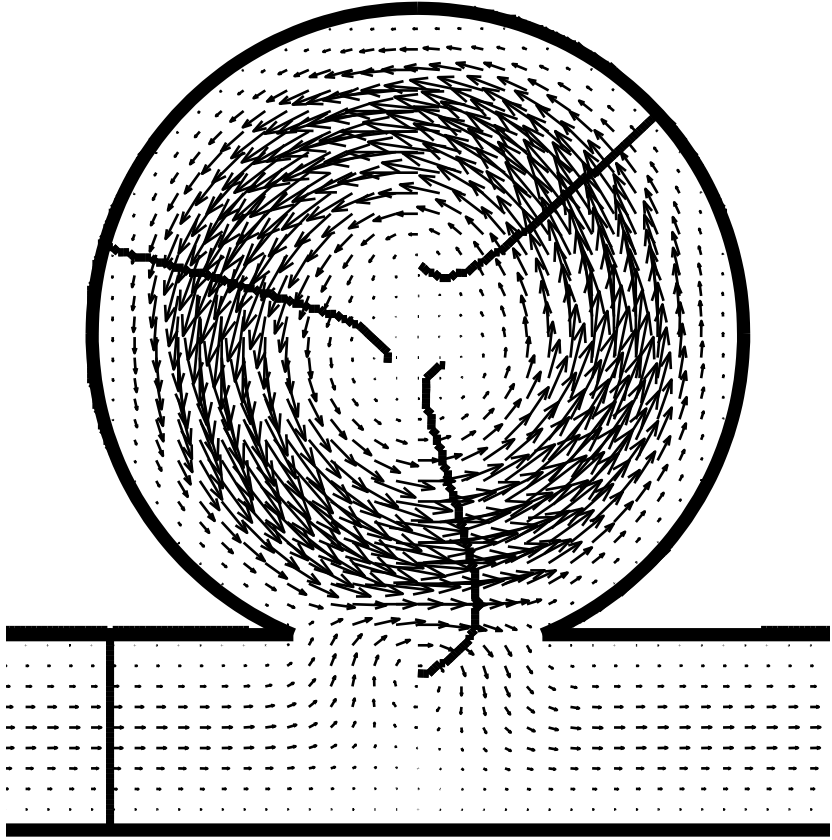
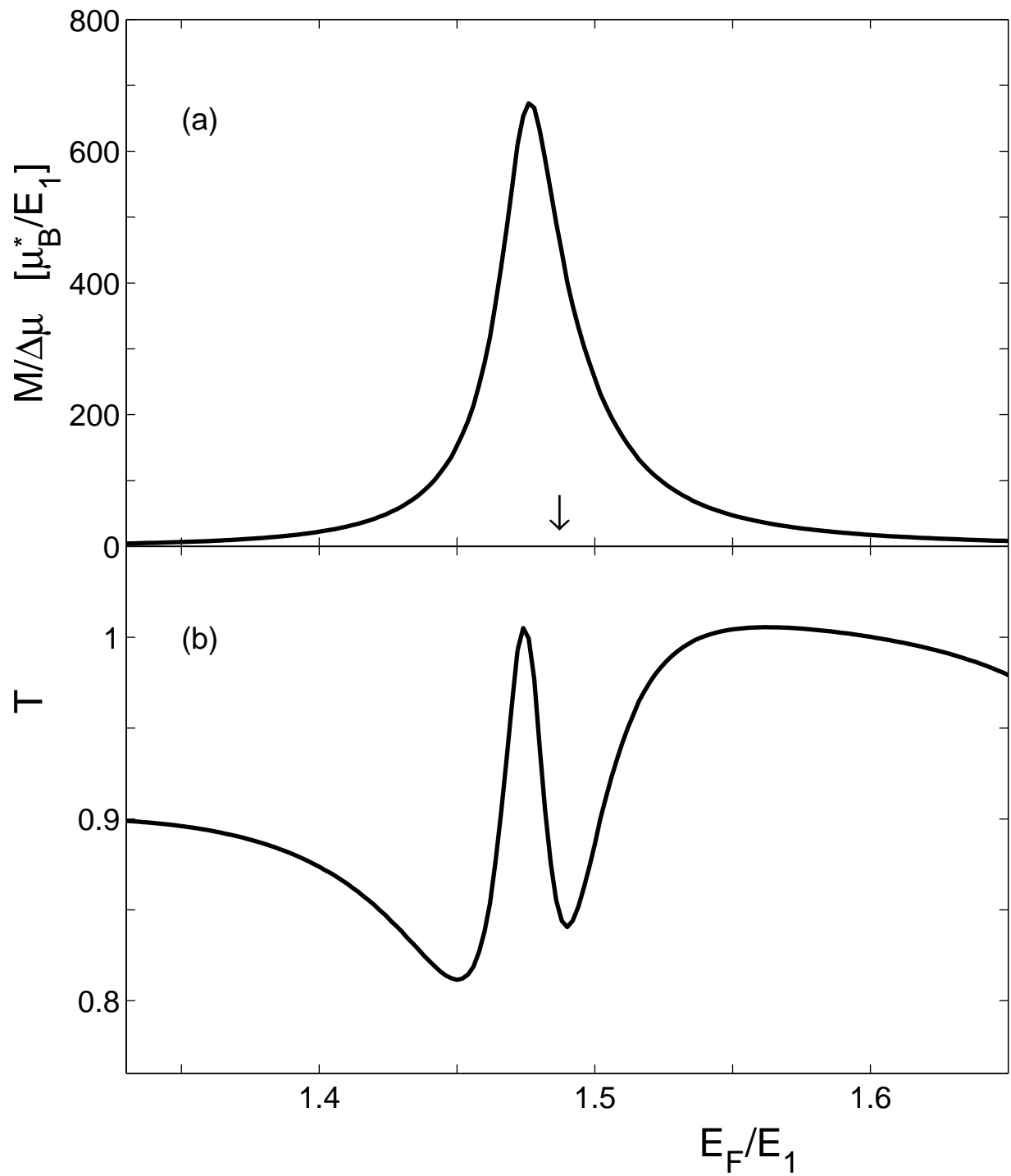
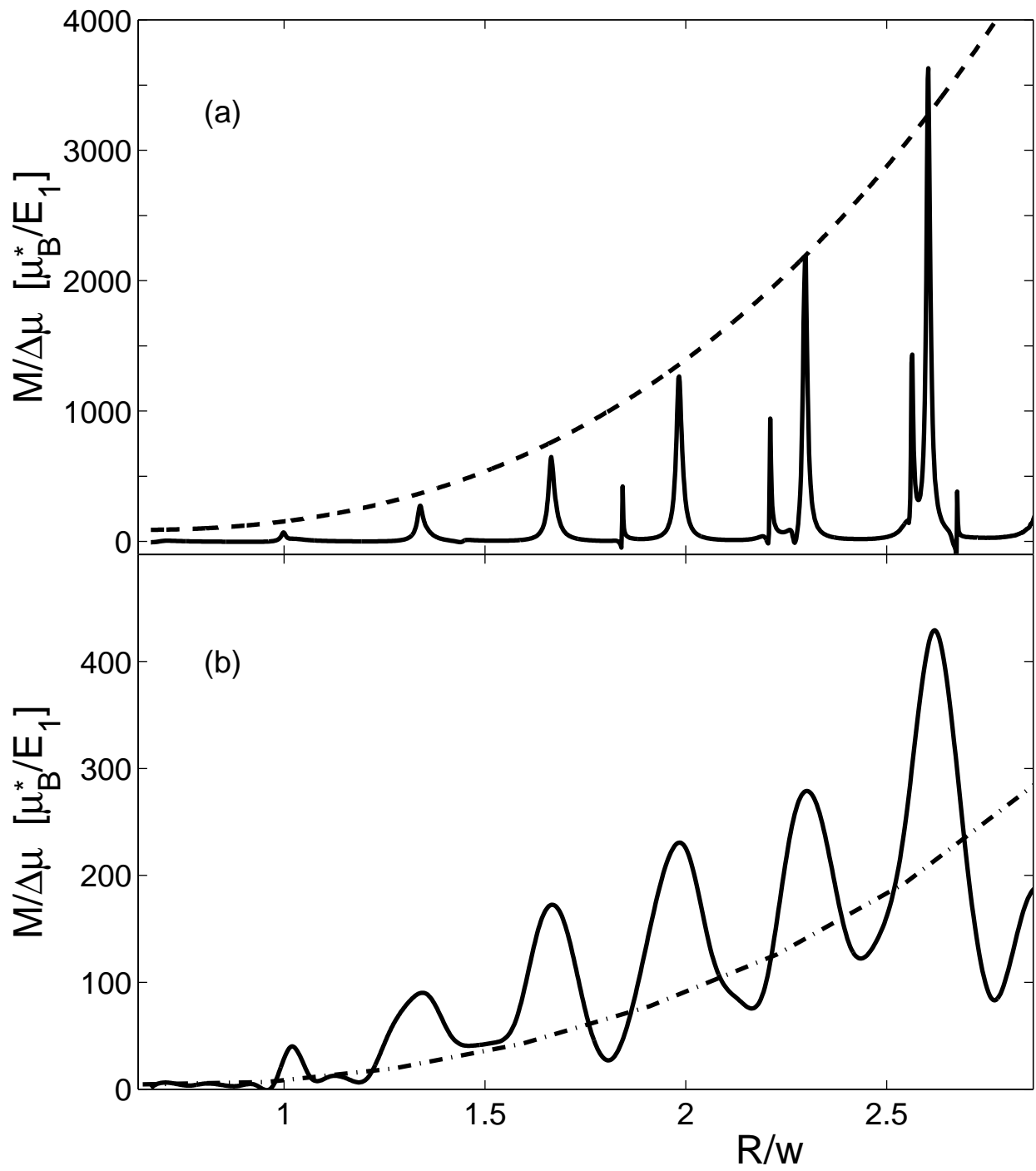
\end{document}